\documentclass[sigconf]{acmart}

\usepackage{multirow}
\usepackage{enumitem}

\AtBeginDocument{%
  }

\setcopyright{acmlicensed}
\copyrightyear{2018}
\acmYear{2018}
\acmDOI{XXXXXXX.XXXXXXX}

\acmConference[Conference acronym 'XX]{Make sure to enter the correct
  conference title from your rights confirmation emai}{June 03--05,
  2018}{Woodstock, NY}
\acmISBN{978-1-4503-XXXX-X/18/06}

\begin{document}

%%
%% The "title" command has an optional parameter,
%% allowing the author to define a "short title" to be used in page headers.
\title{Scaffolding Research Projects in Theory of Computing Courses}

%%
%% The "author" command and its associated commands are used to define
%% the authors and their affiliations.
%% Of note is the shared affiliation of the first two authors, and the
%% "authornote" and "authornotemark" commands
%% used to denote shared contribution to the research.

% \author{Anonymous Author}
% \email{anon@inst.edu}
% % \orcid{0000-0003-1739-1127}
% \affiliation{%
%   \institution{Anonymous Institution}
%   \city{Anonymous City}
%   \state{Anonymous State}
%   \country{Anonymous Country}
%   \postcode{Anonymous Zip}
% }

\author{Ryan E. Dougherty}
\email{ryan.dougherty@westpoint.edu}
\orcid{0000-0003-1739-1127}
\affiliation{%
  \institution{United States Military Academy}
  \streetaddress{601 Thayer Road}
  \city{West Point}
  \state{New York}
  \country{USA}
  \postcode{10996}
}

%%
%% By default, the full list of authors will be used in the page
%% headers. Often, this list is too long, and will overlap
%% other information printed in the page headers. This command allows
%% the author to define a more concise list
%% of authors' names for this purpose.
\renewcommand{\shortauthors}{anonymous author}

%%
%% The abstract is a short summary of the work to be presented in the
%% article.
\begin{abstract}
Theory of Computing (ToC) is an important course in CS curricula because of its connections to other CS courses as a foundation for them. Traditional ToC course grading schemes are mostly exam-based, and sometimes a small weight for traditional proof-type assignments. Recent work experimented with a new type of assignment, namely a ``mock conference'' project wherein students approach and present ToC problems as if they were submitting to a ``real'' CS conference. In this paper we massively scaffold this existing project and provide our experiences in running such a conference in our own ToC course.
\end{abstract}

%%
%% The code below is generated by the tool at http://dl.acm.org/ccs.cfm.
%% Please copy and paste the code instead of the example below.
%%
\begin{CCSXML}

\end{CCSXML}

%%
%% Keywords. The author(s) should pick words that accurately describe
%% the work being presented. Separate the keywords with commas.
\keywords{theoretical computer science, computer science education, large language model, formal languages, automata theory, chatgpt}
%% A "teaser" image appears between the author and affiliation
%% information and the body of the document, and typically spans the
%% page.

\received{20 February 2007}
\received[revised]{12 March 2009}
\received[accepted]{5 June 2009}

%%
%% This command processes the author and affiliation and title
%% information and builds the first part of the formatted document.
\maketitle

\section{Introduction}

In technical Computer Science courses, the goal is to instill proficiency in technical writing, reading, and presenting material to an audience. 
This goal can take several forms, including: writing proofs of claims, devising an algorithm to solve a problem, critiquing others' work in the form of reviews, and giving an oral presentation to an audience of their peers.
A natural avenue for students to take their skills is a CS conference, as it involves all of these forms: proving a theorem, coding an algorithm with experiments, anonymously reviewing other submissions, and (if accepted) presenting the work to other CS educators.
Although there exist many CS conferences, many are out-of-reach for most CS undergraduate students as they lack the technical expertise to both write and read technical CS material effectively, and thus submitted papers and reviews are unlikely to be accepted or considered.
However, an instructor could in principle create a ``mock conference'' in which the experience of participating within a real conference is the same, but without the listed issues above. 

We implemented a mock conference experience called within the theory of computing (ToC) course at our institution based on a first implementation by Dougherty \cite{dougherty2024experiences}. 
This first implementation was unscaffolded, whereas ours is massively scaffolded based on the recommendations Dougherty made.
In this paper we detail our experiences in running such a scaffolded project and give our own recommendations for instructors/educators of larger institutions for pedagogical approaches in implementing such an experience within their technical CS courses.
The rest of this paper is organized as follows:
\begin{itemize}
    \item Section~\ref{sec:related_work} contains related work.
    \item Section~\ref{sec:institutional_context} contains our institutional and course context. 
    \item Section~\ref{sec:design} discusses our course project's design.
    \item Section~\ref{sec:issues} contains all issues and challenges that we faced in running our mock conference.
    \item Section~\ref{sec:reflections} discusses our reflections.
    \item Section~\ref{sec:future_recommend} contains future work items and recommendations for how ToC educators could implement our mock conference project. 
    \item Section~\ref{sec:conclusion} concludes the paper.
\end{itemize}

\section{Related Work}\label{sec:related_work}

Not all ToC courses teach the same content, but a generally applicable background can be found in the first five chapters of \citet{sipser2013introduction}.
Of course, our project scaffolds that of \citet{dougherty2024experiences}.
However, this is the only work that specifically is about research-like experiences in ToC courses; see references there-in for other courses that incorporated some mock conference-like experiences.
Otherwise, ToC related work for CS education is concentrated mainly in visualizations, course design, or auto-grading systems for common problems in large ToC courses. 
Large-scale auto-graded exercises have recently been developed for traditional problems in ToC courses \cite{robert2024fsm,erickson2023auto}.
\citet{xia2023using} used context-free grammars for auto-grading student work in writing precise mathematical statements within an Algorithms course.
\citet{AlurDGKV13} developed an automated grading mechanism for finite automata. 
\citet{mohammed2024teaching} adopted a ``programmed instruction'' approach to teaching ToC, which involved automatically graded exercises from readings.

A popular visual tool for manipulating objects in ToC courses is JFLAP \cite{Rodger06}, but requires local installation and does not support auto-graded exercises.
\citet{chakraborty2011fifty} survey simulators of automata through 2011.  
Of note, \citet{Norton09} developed another extension for JFLAP that tests for finite automata equality, and gives a ``witness'' string if a student submission does not match the model one.
\citet{pereira2018mobile} created a mobile app for ToC course usage.
In 2022 \citet{bezakova2022effective} created an extension to JFLAP with a feedback system for students with incorrect machines, but does not have partial credit.
\citet{marchetti2024redux} developed a visual tool for NP-completeness and mapping reductions; their tool is not applicable to ours as our project does not include NP-Completeness proofs, but the ideas of proof visualization may prove useful for inclusion with ours.

We next give some related works for ToC course and assignment design.
\citet{randolph2024participatory} had an experiment about increasing student control of ToC in-class time. 
\citet{dougherty2024designing} implemented ToC course concepts in a backwards order.
\citet{brunelle2022comfortable} detailed experiences with putting approximately 300 students in large ToC courses into smaller groups for collaboration purposes; there is evidence that such small groups are effective for mathematical topics \cite{webb1991task}.
Our ToC course is much smaller than theirs, but their methods could be partially adapted to ours; see Section~\ref{sec:implement_large_toc} for details.  
\citet{lobo2006np} created a curriculum for teaching NP-Completeness, which is one topic within our ToC course; however, it is not part of our project. 
\citet{maccormick2018strategies} taught non-decision (e.g., optimization, search) problems to their ToC course and found they were preferred by students compared to decision ones.
Their findings are not applicable to our setting as the ToC objects we consider are best well-suited as decision problems.

\section{Institutional and Course Context}\label{sec:institutional_context}
Our institution, an undergraduate-only small liberal-arts college (SLAC) in the northeastern United States, has a ToC course offered once per year with between 30 and 45 CS majors each.
Topics are broken into five major sections: regular languages, context-free languages, Turing Machines and decidability, undecidability, and NP-completeness.
Our ToC course is primarily taken during the students' junior year, with prerequisites being discrete mathematics and digital logic, and a co-requisite of Algorithms.
We had three sections of our ToC course, all taught by the author, with a total of 33 students in 3 sections.
Our study has been approved by our local institutional review board (IRB).
All students who either changed majors, dropped, or elected to be removed from the study have been done so before analysis has taken place.
The IRB has also anonymized the data before the author began analysis.

Following the work of \citet{dougherty2024experiences}, we implemented a scaffolded version of his assignment.
The original version of this assignment created groups of 2 or 3 students each and gave each group a set of problems to solve, all within the regular and context-free language sections. 
The groups had to create a paper within \LaTeX\ and in the style/formatting of a CS conference that solved and discussed those problems.
Further, after the paper submission deadline, students individually and anonymously refereed several papers; our mock conference used a double-blind reviewing scheme.
The due date was approximately 6 weeks after assigning.

Dougherty highlights several major problems with this approach.
First, most proofs within a ToC course happen within the third and fourth course sections, so student mastery of these topics are not assessed in this assignment.
Second, since most students had no research experiences by the time they took ToC, they would have little ability to write a research-style paper. 
To combat these issues, we extended his assignment into a multi-phase project across the semester with scaffolding within each part in hopes to ensure students both took writing the paper seriously and to more closely mimic how a junior researcher would approach a project. 

\section{Project Design}\label{sec:design}

At the beginning of the semester, students were placed in groups of two.
The grouping choice was done by the instructor based on a weighted average of their incoming GPAs in pre-requisite courses, which could be overruled if students had a preferred partner to work with as long as their weighted averages were not significantly different.
If not overruled, students were paired by adjacent averages, even if they were in different ToC sections.
Students could also elect if they wanted to conduct ``real'' publishable research, and if their weighted average was sufficiently high, they were given such a project; two groups (4 students) were in this category.
The instructor at the end of the semester ``accepted'' some of the papers, and those received a small bonus in course grade.

The instructor gave each group a different overall project topic relevant to ToC with some practical application in mind.
For example, one group $G_1$ focused on ``planar'' automata as drawing figures of automata without crossings would be beneficial to ToC instructors and students.
Another group $G_2$ assessed whether generative AI can effectively solve standard ToC course problems.
A third group $G_3$ answered questions about the number of strings of a given length are accepted by various ToC computational models.
The groups and project topics were anonymous to other groups.

\begin{figure*}
    \centering
    \includegraphics[width=0.8\linewidth]{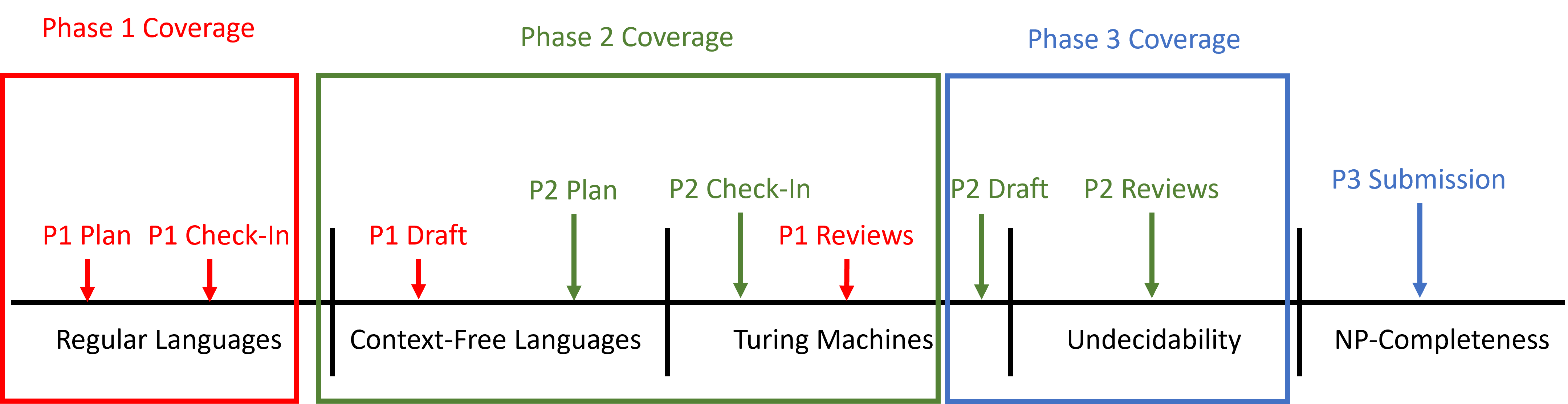}
    \caption{Overview of our scaffolded ToC course research project, including when plans, check-ins, drafts, reviews, and the final submission are due. 
    Section lengths are not drawn exactly to scale.}
    \label{fig:project_overview}
\end{figure*}

The project was worth 45\% of the overall grade, with a breakdown provided below.
Figure~\ref{fig:project_overview} shows a project overview in terms of scheduling; it consists of three phases, each worth 14.5\% of the overall grade.
The instructor gave problems relevant to each phase when the coverage of that phase began. 
For example, for group $G_1$ above the instructor gave questions about DFA and NFA planarity for Phase 1, PDA planarity for Phase 2, and Turing Machine planarity and Undecidability for Phase 3.
For group $G_2$ they were to generate many standard ToC problems; for Phase 1, they focused only on regular language questions, and similar for the other two phases.
Importantly, problems of one phase were not distributed before that phase began to minimize student cognitive load.

All students were given a comprehensive document that contained grading rubrics, writing and reviewing recommendations, and individual phase and submission requirements (both formatting and otherwise).
Students individually took a low-stakes untimed quiz (worth 1.5\% of the overall grade) based on this document that tested whether they have read it and understood the overall project requirements.
All submissions used a standard \LaTeX\ template, and students were given the Overleaf \LaTeX\ tutorial resource\footnote{\url{https://www.overleaf.com/learn/latex/Tutorials}} in case they have not used \LaTeX\ before.  

The first two phases contain four sub-parts: a plan, check-in, draft, and reviews.
We summarize the plan, check-in, and draft sub-parts below.
\begin{itemize}
    \item \textbf{Plan}: a summary of what groups have done so far in researching the problem, and what they plan to do next with regard to the phase's problems. This needs to have some nontrivial amount of work already completed.
    
    \item \textbf{Check-in}: a progress report. Proofs/Figures/etc. need to be started and have some amount of information that indicates that significant progress at solving them has been complete. Otherwise, nontrivial work still has to have been completed at this point.
    \item \textbf{Draft}: a rough draft relevant to the phase's problems. Proofs, figures, etc. must be either fully correct or mostly correct at this point, and the surrounding text should be close to finalized.
    
\end{itemize}

Each plan was worth 1\% of the overall grade, and was a individual submission.
Each check-in was worth 3.5\% of the overall grade, and was a group submission.
Each draft was worth 6.5\% of the overall grade, and was a group submission.
Such submissions would be refereed by other students, and thus were anonymized.

After the phase's drafts were submitted, each student was assigned two papers to anonymously referee; this was worth 3.5\% of the overall grade for each phase.
The template is very similar to ``real'' CS conferences: a brief paper summary, evaluation summary, a written evaluation, and their confidence with regard to the material of the paper.
Students were required to be anonymous and brutally honest in their evaluations, and a list of questions to consider for their reviews, such as missing parts of proofs, professional writing standards, formatting, missing organizational items, etc.
The instructor graded reviews primarily on accuracy and completeness of the reviewer's comments.
Reviews were coordinated through HotCRP\footnote{\url{https://hotcrp.com/}} based on the suggestion in Dougherty \cite{dougherty2024experiences} because EasyChair's\footnote{\url{https://easychair.org}} free version did not support more than 20 total submissions, including deleted ones.

The third and last phase consists of one submission, the final paper, which is analogous to the deadline submission of a ``real'' CS conference.
Students were given more problems relevant to Phase 3, and had to now focus on other aspects of conference paper writing, such as the title, abstract, reference formatting, and professional formatting. 
Phase 3's submission was worth 14.5\% of the course grade (i.e., all three phases overall have equal weight).

\subsection{Example Project Description}
\label{sec:example_project_description}

Here is an example project description with all questions provided in each phase. 

\noindent\textbf{Project Description}: Our course concerns analyzing different types of languages and the properties that they have. Your project will be about analyzing the \emph{lengths} of strings within languages. Let $L$ be a language, and define $length(L)$ to be the set of lengths of strings in $L$. More formally, $length(L) = \{1^{|w|} : w \in L\}$. 

\noindent\textbf{Phase 1}: 
\begin{enumerate}[label=(\alph*)]
    \item Let $M$ be a DFA given to you as input with language $L$. Create an algorithm that will determine whether or not $length(L) = 1^\star$. In other words, determine if $M$ accepts at least one string of every length. 
    \item Same as part (a) but have the algorithm determine instead whether or not $length(L)$ is regular. 
    \item Is there a regular language $L$ for which $length(L)$ is not regular? If so, prove it; otherwise, give a counterexample.
    \item If $length(L)$ is regular, is $L$ necessarily regular? If so, prove it; otherwise, give a counterexample. 
\end{enumerate}

\noindent\textbf{Phase 2}:
\begin{enumerate}[label=(\alph*)]
    \item[(a)-(d)] Same as Phase 1's part (a) through part (d) but instead for context-free languages. 
    \item[(e)] For each of the ``no" answers for Phase 1's (c) and (d), implement an algorithm for the conversion in any programming language of your choice. Generate many different DFAs of varying sizes, perform the conversion, and report the size of the DFA/NFA/etc. that results from the conversion. Then find the minimum-size DFA for that language. Report a comparison between input DFA size and output minimum-DFA size. 
\end{enumerate}

\noindent\textbf{Phase 3}:
\begin{enumerate}[label=(\alph*)]
    \item If $L$ is decidable, then is $length(L)$ decidable? If so, give an algorithm and analyze its runtime; if not, give a proof that it is undecidable. 
    \item If $length(L)$ is decidable, is $L$ decidable? If so, give an algorithm and analyze its runtime; if not, give a proof that it is undecidable. 
    \item Define $REGLEN = \{\langle M\rangle : M$ is a Turing Machine and $length(L(M))$ is regular\}. Is $REGLEN$ decidable? If so, give an algorithm and analyze its runtime; if not, give a proof that it is undecidable.  
\end{enumerate}

\subsection{Course Resources}

As most students have not completed any research prior to our course, here we describe what we gave students for each part of the research process.
Of course, we do not expect students to, on their first attempt, accurately complete a literature review or setup an experimental environment. 
Our goal was to provide general resources without being hyper-specific because (1) group topics varied widely; (2) writing research papers incorporates individual researcher styles, in all meanings; and (3) students should be able to glean some amount of research writing knowledge from reading relevant work.

For performing a short literature review, we recommended students use Google Scholar to find relevant papers based on their problems.
We also provided instructions on how to accurately cite such papers.
% , both for using \BibTeX{} and for in-text citations.
As ToC is a mathematically inclined course, we gave advice on how to use the built-in \LaTeX\ environments for theorems, lemmas, and proofs, as they would be relevant to any group (dis)proving statements.
Students were given a lengthy description on how to use such environments.

A large number of groups performed an experimental evaluation, usually involving a mathematical conversion or construction they prove in their paper.
Then they implement it, analyze it from a big-O perspective, and run experiments on the time taken; directions on how to do these were also provided.
One purpose was to draw connections with the Algorithms co-requisite course.
Students were encouraged to think about input/output format, data structure choice, big-O analysis, programming language choice, and experimental setup. 
For all such groups, the instructor gave general instructions on how to construct appropriate inputs for an experiment; usually this involves generating many different automata with hundreds or thousands of states and transitions to notice trends when considering runtime. 
A link was provided to an existing automata library called {\tt automata-lib} \cite{Evans_automata_A_Python_2023} for group use provided it was cited correctly.

\section{Issues and Challenges}\label{sec:issues}

We believe that for specifically this project and course structure, it is essential to have a small number of students when implementing our project.
The main reasons for this belief are due to administrative course overhead.
The first major overhead was the time spent generating the project topics as they needed to be distinct from other groups; Dougherty \cite{dougherty2024experiences} had the same issue.
The second major overhead was grading; even though there were ``only'' 17 groups, each of them produced three plans, check-ins, and drafts, and one final research paper. 
There was only one instructor, and thus the grading load was immense.
Further still, each student produced two full-length reviews of other papers. 
Each submission needed to have detailed feedback, similar to how a junior researcher would submit a draft paper.
However, there are potential avenues for which this project can be used in very large courses provided there is sufficient TA support; see Section~\ref{sec:implement_large_toc} for details.

\section{Reflections}\label{sec:reflections}

To identify patterns across the three phases and how they impact the rest of our ToC course, grades for our project are given in Table~\ref{tbl:grades}.
Groups are sorted according to the overall score of the final paper submission (Phase 3) and group numbers are listed based on this ordering.
We did not conduct a survey, but collected (anonymously made) student IDs and individual/group grades. 

% Purple = A+
% Light blue = A
% Blue = A-
% Light green = B+
% Green = B
% Dark Green = B-
% Light yellow = C+
% Yellow = C
% Dark Yellow = C-
% Orange = D
% Red = F
\begin{table*}[]
\caption{\label{tbl:grades}Draft, Plan + Check-In + Draft, and Overall Grades for Phases 1 and 2, Correctness and Overall grade for Phase 3, an average of all phases for Plan + Check-In + Draft, the same but also including reviews, and Final Exam grades for all groups in our ToC course. All entries are averaged across the group members and normalized as percentages.
Groups are ordered by score in the ``Overall'' for Phase 3 row.
The color coding is based on the letter grades for our institution: light blue is an A, dark blue is A-, light green is B+, green is B, dark green is B-, light yellow is C+, yellow is C, dark yellow is C-, orange is D, and red is F. 
}

\begin{tabular}{c|l|l|l|l|l|l|l|l|l|l|l|l|l|l|l|l|l|l|}
\cline{2-19}
\multicolumn{1}{l|}{}                                                                        & Group \#   & \multicolumn{1}{c|}{1}       & \multicolumn{1}{c|}{2}       & \multicolumn{1}{c|}{3}       & \multicolumn{1}{c|}{4}       & \multicolumn{1}{c|}{5}       & \multicolumn{1}{c|}{6}       & \multicolumn{1}{c|}{7}       & \multicolumn{1}{c|}{8}       & \multicolumn{1}{c|}{9}       & \multicolumn{1}{c|}{10}      & \multicolumn{1}{c|}{11}      & \multicolumn{1}{c|}{12}      & \multicolumn{1}{c|}{13}      & \multicolumn{1}{c|}{14}      & \multicolumn{1}{c|}{15}      & \multicolumn{1}{c|}{16}      & \multicolumn{1}{c|}{17}      \\ \hline
\multicolumn{1}{|c|}{}                                                                       & Draft      & \cellcolor[HTML]{34CDF9}90.0 & \cellcolor[HTML]{34FF34}83.8 & \cellcolor[HTML]{F8FF00}76.9 & \cellcolor[HTML]{38FFF8}93.8 & \cellcolor[HTML]{FFFFC7}77.7 & \cellcolor[HTML]{9AFF99}89.2 & \cellcolor[HTML]{34FF34}83.8 & \cellcolor[HTML]{FE0000}58.5 & \cellcolor[HTML]{D1D700}71.5 & \cellcolor[HTML]{F8FF00}76.2 & \cellcolor[HTML]{FE0000}52.3 & \cellcolor[HTML]{FE0000}61.5 & \cellcolor[HTML]{9AFF99}88.5 & \cellcolor[HTML]{F8A102}67.7 & \cellcolor[HTML]{F8A102}66.2 & \cellcolor[HTML]{F8A102}67.7 & \cellcolor[HTML]{FE0000}58.5 \\ \cline{2-19} 
\multicolumn{1}{|c|}{}                                                                       & P+C+D      & \cellcolor[HTML]{34FF34}85.9 & \cellcolor[HTML]{FFFFC7}78.6 & \cellcolor[HTML]{32CB00}81.4 & \cellcolor[HTML]{34CDF9}91.8 & \cellcolor[HTML]{32CB00}81.4 & \cellcolor[HTML]{32CB00}81.4 & \cellcolor[HTML]{32CB00}82.7 & \cellcolor[HTML]{FE0000}63.6 & \cellcolor[HTML]{F8FF00}76.4 & \cellcolor[HTML]{FFFFC7}77.3 & \cellcolor[HTML]{FE0000}60.9 & \cellcolor[HTML]{F8A102}68.2 & \cellcolor[HTML]{34FF34}85.0 & \cellcolor[HTML]{D1D700}71.8 & \cellcolor[HTML]{F8A102}67.3 & \cellcolor[HTML]{D1D700}70.0 & \cellcolor[HTML]{F8A102}66.4 \\ \cline{2-19} 
\multicolumn{1}{|c|}{\multirow{-3}{*}{\begin{tabular}[c]{@{}c@{}}Phase \\ \#1\end{tabular}}} & Overall    & \cellcolor[HTML]{34CDF9}90.0 & \cellcolor[HTML]{34FF34}84.5 & \cellcolor[HTML]{32CB00}82.8 & \cellcolor[HTML]{38FFF8}94.5 & \cellcolor[HTML]{34FF34}86.6 & \cellcolor[HTML]{34FF34}84.8 & \cellcolor[HTML]{34FF34}85.5 & \cellcolor[HTML]{F8A102}69.0 & \cellcolor[HTML]{32CB00}80.7 & \cellcolor[HTML]{32CB00}81.4 & \cellcolor[HTML]{F8A102}67.6 & \cellcolor[HTML]{F8FF00}74.8 & \cellcolor[HTML]{34FF34}86.6 & \cellcolor[HTML]{FFFFC7}77.2 & \cellcolor[HTML]{D1D700}70.0 & \cellcolor[HTML]{D1D700}70.7 & \cellcolor[HTML]{F8FF00}73.8 \\ \hline
\multicolumn{1}{|c|}{}                                                                       & Draft      & \cellcolor[HTML]{D1D700}72.3 & \cellcolor[HTML]{34CDF9}90.0 & \cellcolor[HTML]{D1D700}72.3 & \cellcolor[HTML]{34CDF9}91.5 & \cellcolor[HTML]{34FF34}83.1 & \cellcolor[HTML]{34FF34}85.4 & \cellcolor[HTML]{9AFF99}88.5 & \cellcolor[HTML]{F8FF00}76.2 & \cellcolor[HTML]{FFFFC7}77.7 & \cellcolor[HTML]{D1D700}70.0 & \cellcolor[HTML]{FFFFC7}77.7 & \cellcolor[HTML]{F8FF00}76.2 & \cellcolor[HTML]{32CB00}81.5 & \cellcolor[HTML]{D1D700}70.0 & \cellcolor[HTML]{F8FF00}73.8 & \cellcolor[HTML]{FE0000}62.3 & \cellcolor[HTML]{F8A102}68.5 \\ \cline{2-19} 
\multicolumn{1}{|c|}{}                                                                       & P+C+D      & \cellcolor[HTML]{D1D700}70.0 & \cellcolor[HTML]{9AFF99}89.5 & \cellcolor[HTML]{F8FF00}76.8 & \cellcolor[HTML]{9AFF99}88.6 & \cellcolor[HTML]{34FF34}83.1 & \cellcolor[HTML]{34CDF9}90.9 & \cellcolor[HTML]{34CDF9}90.0 & \cellcolor[HTML]{FFFFC7}79.1 & \cellcolor[HTML]{FFFFC7}79.1 & \cellcolor[HTML]{32CB00}80.0 & \cellcolor[HTML]{32CB00}81.4 & \cellcolor[HTML]{32CB00}80.9 & \cellcolor[HTML]{F8FF00}76.4 & \cellcolor[HTML]{F8FF00}73.6 & \cellcolor[HTML]{FFFFC7}78.2 & \cellcolor[HTML]{F8A102}68.2 & \cellcolor[HTML]{F8FF00}74.1 \\ \cline{2-19} 
\multicolumn{1}{|c|}{\multirow{-3}{*}{\begin{tabular}[c]{@{}c@{}}Phase \\ \#2\end{tabular}}} & Overall    & \cellcolor[HTML]{F8FF00}76.7 & \cellcolor[HTML]{34CDF9}92.8 & \cellcolor[HTML]{32CB00}81.7 & \cellcolor[HTML]{34CDF9}92.1 & \cellcolor[HTML]{9AFF99}87.9 & \cellcolor[HTML]{38FFF8}93.1 & \cellcolor[HTML]{34CDF9}91.9 & \cellcolor[HTML]{32CB00}80.0 & \cellcolor[HTML]{34FF34}83.4 & \cellcolor[HTML]{34FF34}84.1 & \cellcolor[HTML]{34FF34}86.6 & \cellcolor[HTML]{34FF34}86.2 & \cellcolor[HTML]{FFFFC7}78.6 & \cellcolor[HTML]{FFFFC7}77.1 & \cellcolor[HTML]{32CB00}82.8 & \cellcolor[HTML]{F8FF00}76.6 & \cellcolor[HTML]{FFFC9E}78.3 \\ \hline
\multicolumn{1}{|c|}{}                                                                       & Accuracy   & \cellcolor[HTML]{9AFF99}88.6 & \cellcolor[HTML]{F8FF00}74.2 & \cellcolor[HTML]{34FF34}83.6 & \cellcolor[HTML]{34FF34}83.3 & \cellcolor[HTML]{32CB00}82.2 & \cellcolor[HTML]{F8A102}67.8 & \cellcolor[HTML]{D1D700}71.2 & \cellcolor[HTML]{F8A102}66.3 & \cellcolor[HTML]{FE0000}64.5 & \cellcolor[HTML]{FE0000}51.7 & \cellcolor[HTML]{D1D700}70.3 & \cellcolor[HTML]{FE0000}64.6 & \cellcolor[HTML]{FE0000}61.1 & \cellcolor[HTML]{FE0000}57.6 & \cellcolor[HTML]{FE0000}70.9 & \cellcolor[HTML]{FE0000}58.9 & \cellcolor[HTML]{FE0000}48.6 \\ \cline{2-19} 
\multicolumn{1}{|c|}{\multirow{-2}{*}{\begin{tabular}[c]{@{}c@{}}Phase\\ \#3\end{tabular}}}  & Overall    & \cellcolor[HTML]{38FFF8}93.7 & \cellcolor[HTML]{34CDF9}92.6 & \cellcolor[HTML]{34CDF9}90.1 & \cellcolor[HTML]{9AFF99}89.3 & \cellcolor[HTML]{34FF34}85.4 & \cellcolor[HTML]{32CB00}81.5 & \cellcolor[HTML]{F8FF00}75.8 & \cellcolor[HTML]{F8A102}67.8 & \cellcolor[HTML]{F8A102}66.7 & \cellcolor[HTML]{F8A102}66.6 & \cellcolor[HTML]{F8A102}65.0 & \cellcolor[HTML]{FE0000}64.7 & \cellcolor[HTML]{FE0000}63.2 & \cellcolor[HTML]{FE0000}63.1 & \cellcolor[HTML]{FE0000}61.4 & \cellcolor[HTML]{FE0000}58.8 & \cellcolor[HTML]{FE0000}54.6 \\ \hline
\multicolumn{1}{|c|}{}                                                                       & P+C+D      & \cellcolor[HTML]{34FF34}85.3 & \cellcolor[HTML]{9AFF99}88.8 & \cellcolor[HTML]{FFFFC7}79.8 & \cellcolor[HTML]{34CDF9}91.6 & \cellcolor[HTML]{32CB00}82.1 & \cellcolor[HTML]{34FF34}85.4 & \cellcolor[HTML]{32CB00}82.7 & \cellcolor[HTML]{F8A102}67.5 & \cellcolor[HTML]{D1D700}72.0 & \cellcolor[HTML]{D1D700}70.9 & \cellcolor[HTML]{F8A102}65.0 & \cellcolor[HTML]{F8A102}67.4 & \cellcolor[HTML]{FFFFC7}77.7 & \cellcolor[HTML]{F8A102}66.9 & \cellcolor[HTML]{F8A102}67.1 & \cellcolor[HTML]{FE0000}62.9 & \cellcolor[HTML]{FE0000}60.5 \\ \cline{2-19} 
\multicolumn{1}{|c|}{\multirow{-2}{*}{All}}                                                  & Overall    & \cellcolor[HTML]{34FF34}86.8 & \cellcolor[HTML]{9AFF99}89.9 & \cellcolor[HTML]{34FF34}84.9 & \cellcolor[HTML]{34CDF9}92.0 & \cellcolor[HTML]{34FF34}86.6 & \cellcolor[HTML]{34FF34}86.5 & \cellcolor[HTML]{34FF34}84.4 & \cellcolor[HTML]{D1D700}72.2 & \cellcolor[HTML]{F8FF00}76.9 & \cellcolor[HTML]{FFFFC7}77.4 & \cellcolor[HTML]{F8FF00}73.0 & \cellcolor[HTML]{F8FF00}75.2 & \cellcolor[HTML]{F8FF00}76.1 & \cellcolor[HTML]{D1D700}72.5 & \cellcolor[HTML]{D1D700}71.4 & \cellcolor[HTML]{F8A102}68.7 & \cellcolor[HTML]{F8A102}68.9 \\ \hline
\multicolumn{1}{l|}{}                                                                        & Final Exam & \cellcolor[HTML]{32CB00}82.1 & \cellcolor[HTML]{34FF34}86.9 & \cellcolor[HTML]{34FF34}84.6 & \cellcolor[HTML]{9AFF99}87.4 & \cellcolor[HTML]{9AFF99}87.7 & \cellcolor[HTML]{32CB00}81.5 & \cellcolor[HTML]{34FF34}85.2 & \cellcolor[HTML]{FE0000}60.7 & \cellcolor[HTML]{F8A102}69.3 & \cellcolor[HTML]{D1D700}70.2 & \cellcolor[HTML]{F8A102}66.4 & \cellcolor[HTML]{FE0000}58.1 & \cellcolor[HTML]{9AFF99}88.0 & \cellcolor[HTML]{34FF34}83.9 & \cellcolor[HTML]{F8FF00}73.4 & \cellcolor[HTML]{FE0000}62.8 & \cellcolor[HTML]{F8A102}69.6 \\ \cline{2-19} 
\end{tabular}
\end{table*}

% grouped r^2 all phases vs final exam = 0.61
% individual r^2 all phases vs final exam ind = 0.46

% wpr1 vs phase1 draft r^2 = 0.49
% wpr2 vs phase2 draft r^2 = 0.22

\subsection{What Worked Well}

There were no late or missing draft or final paper submissions, thanks to the scaffolding. 
Since there were more intermediate deadlines, students could now focus on smaller tasks before they needed to think of the paper in a ``big picture'' context. 
The comprehensive document also helped with students as they did not need to consult the instructor very often for assistance on \LaTeX\ or word choice in writing sections.

Dougherty \cite{dougherty2024experiences} had an issue with EasyChair as it has a limited number of total submissions, even deleted ones, for their free version.
Our usage of HotCRP significantly helped in this respect as there was no limitation on reviews or submissions. 
All administratively relevant features that EasyChair has are also available on HotCRP. 

Writing research-style ToC-type papers (at any level) requires a perspective of the problem domain across different formally defined objects.
It was therefore advantageous to have a single theme for each project as groups could now write an abstract, and intro and concluding sections more effectively than if there was a single phase. 
Students had a smaller learning curve about the general project topic in later phases for this reason.

Having the project cover nearly all topics gave more insights into where and why students struggle, especially with proof writing.
Additionally, students took three exams so comparisons of mistakes can be made.
Generally, students made the same mistakes in their papers as their exams: incorrect assumptions (e.g., being able to apply the Pumping Lemma at all) and forgetting required cases (e.g., induction hypothesis).
Interestingly, there was an essence of perceived professionalism: even if a proof is not correct, the text there was written \emph{as if} it were correct through mathematical language and phrasing. 
Only a very close inspection would reveal any technical inaccuracy.
A potential cause may be that their peers would be directly viewing and critiquing their work.
Hopefully this will inspire future work to understand precisely how students view and fact-check technical content.

% The requirement for most groups to implement some conversions and/or algorithms, and to analyze them, was very beneficial from a pedagogical perspective. 
% More here...

\subsection{What Did Not Work Well}

The comprehensive document was overwhelming for some students as it contained all instructions, including for later phases. 
There were several instances among low-performing students that they did not know they needed to turn in a plan or check-in document, even though they accurately answered the initial quiz's questions about when those items were due and what was required.

A simple $t$-test yields no statistical significance between the ``All Phases'' rows of Table~\ref{tbl:grades} and the overall paper scores of Dougherty \cite{dougherty2024experiences}, despite the scaffolding in our implementation.
This result may be due to a lower-performing population; the incoming major GPA for our offering compared to the previous offering differed by only 0.01, but the outgoing GPA was 0.35 lower in our offering (2.56 vs. 2.91).
Even though the incoming average GPAs were virtually the same, their ability to perform in technical and mathematical courses may have been much lower.
Anecdotally, students had much more trouble with concepts of pre-requisite material than usual, and some even had issue with material in pre-requisites \emph{of} pre-requisites.
However, more sophisticated statistical tests need to be performed, such as mixed-methods \cite{ebling2024analogies}.

Many groups had trouble managing their document's complexity within Phase 3 due to their length. 
Some had submissions wherein they did not modify any, or very little, text from Phases 1 and 2 even after feedback was provided. 
Take note of the large number of red entries in Phase 3's rows.
Groups largely had lower performance in terms of accuracy with Phase 3, partially due to the material notoriously having the most difficult proofs, and also due to integrating all three phases together.
This is no surprise due to the amount of research dedicated to pedagogy of undecidable proofs in the ToC classroom \cite{blumenthal2022teach,ullas2022if,delvado2021learning,delvado2024introducing,dougherty2024designing}.
The fact that such groups largely did not fix earlier issues combined with this explains why there are so many red entries.

Although uncommon, not every group's problems covered the exact same topics.
For example, a correct solution to the problems in the description in Section~\ref{sec:example_project_description} above does not include a Pumping Lemma proof. 
See Section~\ref{sec:implement_large_toc} for a potential automated remedy to this issue.

The provided rubrics may have changed how some groups approached paper writing as they only needed to include what was asked.
In the author's experience, undergraduate CS students crave structure.
Although research papers from a high-level perspective do have structure (intro, related work, results, etc.), the low-level word, image, table, etc. choices distinguish one paper from another.

Consider the last three rows of Table~\ref{tbl:grades}: the average of all phases for each group for the plans, check-ins, and drafts (P+C+D); the same but incorporating reviews (Overall); and the average final exam score for the same group.
There was a fairly high correlation ($r = 0.61$) between the final exam and P+C+D, and a smaller correlation ($r=0.52$) between the final exam and Overall, which gives an indication that this project can somewhat accurately predict overall class performance on the final exam. 
% [TODO FIX THIS] 
However, when one instead considers the individual vs. non-grouped scores, the correlation is somewhat lower (Overall: $r = 0.44$).
This hints at some of the groups having one member's being significantly better at the material than the other at the semester's end despite the instructor's attempts to curb this at the beginning of the semester.
For example, groups 8 and 12 scored (on average) more than one letter grade lower on the final exam compared to their project scores. 
In both cases, one group member scored much higher on the final than the other did, approximately 16\% for both.
More analysis is needed to determine which students would perform more strongly at the end of ToC based on early semester data.

We wanted to assess whether or not this project had any impact on students' abilities to present technical material, even if it was unrelated to their project.
We found almost zero correlation between students' overall performance on our project and their NP-Completeness oral presentation overall scores ($r^2 = 0.004$).
Groups for the presentation were generated randomly within each section.
Most students for this presentation were in different groups than for the project, which may account for why there is no correlation.
Based on the second and third-to-last rows in Table~\ref{tbl:grades}, we are uncertain as to whether a topic shift from NP-completeness to their own projects would make much of a difference in correlation.

\subsection{Limitations}

Like \cite{dougherty2024experiences}, there was no time at the end of the semester for group formal presentations of their papers, and thus students missed out on a central piece of the ``real'' conference experience. 

Although we eliminated bias in assigning students to groups, there still may be instructor bias, from assigning topics to groups and in creating the rubrics; this was also noted by \citet{dougherty2024experiences}.
There was only one instructor, and therefore grading choices or acceptance to the conference may have been affected by a potential bias.

\subsection{Implementation for Large ToC Courses}
\label{sec:implement_large_toc}

We have a potential idea that may be adopted for large ToC offerings that significantly lowers grading and administrative overhead, but does require TA support and coordination.\footnote{Note that this idea is specifically for ToC as there are automatically gradable conditions for some objects. Adopting this idea to other technical CS courses, such as Algorithms or Operating Systems, is possible but changes in how problems/topics are generated and overall structure need to be made.}
The two main factors are (1) the initial generation of project topics so that groups have unique topics, and (2) the time needed to provide feedback for paper submissions. 
To accomplish (1), generate ``regular-type'' conditions for strings for each of the groups.
This process can be automated; one example is to first choose two random distinct strings $s, t$ over an alphabet such as $\{a, b, c\}$.
Then for Phase 1, one can generate the following problems for each group (of course, not limited to these):
\begin{enumerate}
    \item Create an NFA that accepts the language of all strings that do contain $s$ as a substring.
    \item Create a regex for all strings that do NOT contain $s$ as a substring.
    \item Design and analyze an algorithm that, given a DFA $D$, determines if $D$ accepts any string that contains $s$ as a substring.
    \item Same as the previous problem, but determine if $D$ accepts only a finite number of strings that contain $s$ as a substring.
    \item Consider the language $L$ of all strings that have an equal number of $s$ and $t$ substrings.
    Is $L$ regular? 
\end{enumerate}
One can substitute ``substring'' with other properties, such as ``subsequence'', ``prefix'', etc. 
For problem (5) if the instructor's goals include exercising a proof that $L$ is not regular, one needs to have $s, t$ satisfying a certain condition, which is known explicitly \cite{colbourn2018counting} and can also be automated.
For Phase 2, one can use similar ideas but for context-free grammars and pushdown automata instead, and/or for students to implement (1) through (4) in a given program with a set structure that can be automatically graded.
By this, we mean required function stubs that students fill in.
If the institution has a high-performance computing center, one could automatically provide runtime data for (3) and (4) to students to include in their paper.
This would allow students to still take part in interpreting and presenting their data. 
For Phase 3, problems that use undecidability and Rice's theorem can easily be generated.

Some correctness parts can be automatically graded for Phase 1; for Phases 2 and 3 however, there are classic results that equality testing for context-free and Turing Machine models is undecidable \cite{sipser2013introduction}.
One could in principle use a testing method up to a given length/transition limit \cite{bansal2013jflap}.

There still remains feedback and grading aspects.
If students are required to submit their \LaTeX\ files based on a provided template with required paper section stubs, then an automated script can extract these individual sections to distribute to TAs for grading consistency.
% The instructor should make clear what needs to be included for Phase 2, specifically whether the Phase 1 text should also be included/modified.
Use of large language models (LLMs) is currently limited in verifying correctness of ToC mathematical statements \cite{golesteanu2024can} but future LLMs could in the future assist for grading student ToC papers.
% HotCRP is a great tool for handling many submissions.
In the absence of automated paper grading tools, one could still further simulate the administrative side of a CS conference.
Students can take the role of reviewers, TAs can act as meta-reviewers, and the instructor(s) act as conference chairs, providing final decisions.
There is evidence from data science courses that such a process is effective \cite{bhavya2021scaling}.
We recommend that instructors institute a page limit (like ``real'' conferences) to minimize student and TA workload.

We should note that the spirit of this assignment should not be lost in large ToC offerings, as too much automation can potentially stifle student creativity (e.g., standard \LaTeX\ template, figure/table format).
Our offering had two groups that chose to perform ``real'' research in this project, and thus an entirely automated method would remove that possibility.  
One possible avenue to make this project more collaborative is the method of \citet{brunelle2022comfortable} by partitioning a large ToC course into smaller groups; perhaps an ungraded peer evaluation session wherein students skim each others' work and provide quick, general feedback.

\section{Future Work and Recommendations}\label{sec:future_recommend}

For future work, we plan on continuing this project in future ToC offerings. 
We are currently undergoing statistical analyses, both in whether this project is effective (statistically) and how much of an impact it has on student performance in later courses in our CS curriculum.

Observe the columns of Table~\ref{tbl:grades}; groups 1 through 7 appear to have much higher grades than groups 8 through 17, with occasional exceptions such as some entries for group 13 and the final exam score for group 14. 
Based on the color coding, there appears to be an overall bimodal distribution.
Although there does not seem to be related work about bimodality in grade distributions for ToC courses, Patitsas et al. \cite{patitsas2019evidence} gave evidence that bimodality does not exist for introductory CS courses; note that their course was far larger than ours.
Like their work we applied Hartigan's dip test to each of the rows in Table~\ref{tbl:grades}, but only a very small percentage of the samples tested positive for bimodality.
Future work should further investigate how much bimodality exists, if any.

We next discuss how future ToC instructors can even more effectively implement our assignment. 
Of course one needs to consider whether such a project would be effective based on institutional and course context, in particular determining student pre-requisite knowledge of technical writing, course size, topic ordering, and TA support.
If the course is large, we recommend instituting a plagiarism checker and automated grading mechanisms like those described in Section~\ref{sec:implement_large_toc}.
We also recommend providing example past anonymized student submission (if any) or some recommended existing papers to help groups get started. 

We of course recommend incorporating presentations to complete the conference experience.
However, any moderately large ToC course will have time issues in scheduling presentations for each group; we do not recommend making groups larger.
Based on experiences surrounding the COVID-19 pandemic and academic conferences, we recommend ToC instructors consider other modalities of presentation, such as pre-recorded video of a group presentation about their paper.

\section{Conclusion}\label{sec:conclusion}

In this paper we provided our experiences in implementing and running a scaffolded version of a Theory of Computing mock conference project.
We determined that this scaffolding provides more insight into student performance and more closely resembles ``real'' research paper writing experiences. 
Additionally, giving students multiple attempts along with many resources limited last-minute student panic for when deadlines arrived.
Our results determined that there is a rather strong correlation between paper writing and final exam scores, signifying that this is a suitable project for ToC courses.
However, our project was limited due to administrative and grading overhead; to potentially avoid this limitation, we provided ideas for which this project can be generalized to larger ToC offerings.

%%
%% The acknowledgments section is defined using the "acks" environment
%% (and NOT an unnumbered section). This ensures the proper
%% identification of the section in the article metadata, and the
%% consistent spelling of the heading.
\begin{acks}
The opinions in this work are solely of the author, and do not necessarily reflect those of the U.S. Army, U.S. Army Research Labs, the U.S. Military Academy, or the Department of Defense.
\end{acks}

%%
%% The next two lines define the bibliography style to be used, and
%% the bibliography file.
\bibliographystyle{ACM-Reference-Format}
\bibliography{sample-base}

\end{document}